\documentclass[
aip, 
jmp,
amsmath, amssymb, 
reprint,
]{revtex4-1}

\usepackage{graphicx}
\usepackage{makeidx}
\usepackage{multirow}
\usepackage{amssymb}
\usepackage{amsfonts}
\usepackage{amsmath}
\usepackage{amsthm}
\usepackage{natbib}

\usepackage{graphicx}
\usepackage{dcolumn}
\usepackage{bm}

\usepackage[utf8]{inputenc}
\usepackage[T1]{fontenc}
\usepackage{mathptmx}
\usepackage{amsfonts}
\usepackage{bbm}
\usepackage[bottom]{footmisc}

\begin{document}

\preprint{}

\title{On the Bell experiments; an epistemological approach}

\author{Inge S. Helland}
 \email{ ingeh@math.uio.no}
\affiliation{Department of Mathematics, University of Oslo \\ P.O.Box 1053 Blindern, N-0316 Oslo, Norway
\\ Orcid: 0000-0002-7136-873X}


\begin{abstract}
The Nobel prize in physics for 2022 was given for performing Bell experiments with varying degree of sophistication. The interpretation of this experiment is discussed by first recalling Bell's simple argument behind his inequalities, in particular the CHSH inequality. It is argued that any independent observer must have a limitation: He is not able to keep all relevant variables in his mind at the same time when trying to model the experiment. This is contrasted to the solution proposed by Tim Maudlin, where nonlocality is crucial.Maudlin criticizes the Nobel price committee, but this critique is countered. Finally, a related new approach to the foundation of quantum mechanics is briefly sketched, giving references to the relevant literature.

\end{abstract}

\maketitle

\bigskip

\section{Introduction}
\label{Sec1}

The Nobel prize in physics for 2022 has been given to three experimental physicists for having performed, with various degrees of sophistication, a so-called Bell experiment. This experiment has showed beyond any doubt that what is called the CHSH inequality, one special among several Bell inequalities, may be violated in practice. This agrees with the predictions of quantum mechanics but is not in agreement with what may be called a simple common- sense argument, an argument given originally by the theoretical physicist John Bell, whose fundamental works in the 1960’s raised questions that have led to many different articles since then in the physical literature.
Indeed, several physicists and philosophers have tried to understand the experimental result that now has been made 100 \% clear. My own understanding is given in the recent article [1]. It is based upon a simple model of an observer’s mind, a model from which essential elements of quantum mechanics may be deduced [2,3]. This whole approach implies a totally new way of looking upon quantum theory, perhaps a way which may be of some interest now in connection to the results of the Bell experiments.

Other researchers have other ways of understanding the results of the Bell experiment. I can for instance refer to the recent article by the philosopher Tim Maudlin [4]. Maudlin claims that the Physics Nobel Prize is misunderstood even by the Nobel prize committee itself. I must disagree with Maudlin here. Of course, he is right in much of which he says in his article, but in my opinion the Nobel prize committee was also right in what it said. To explain my points of view, I must take as a point of departure, first the original Bell arguments around his inequalities, as modified to the CHSH inequality, then my own simple model of an observer’s (or in fact any person’s) mind.

\section{The Bell argument}
\label{Sec2}

Here are the arguments essentially due to John Bell: Assume that a person C is trying to make sense of the Bell experiment, and in his mind, he focuses on just one run of the experiment, an experiment where particles with spin are sent to two actors Alice and Bob. Here Alice makes a choice between two settings of for measuring a spin component: a (giving response, say A) or a' (response A'). Similarly, Bob makes a choice between settings b (hypothetical response B) or b' (hypothetical response B'). Without loss of generality, the responses can be taken to be +1 or -1 (spin up or down, or left or right, or whatever).

Call this person Charlie. Assume that - and that is the crucial assumption - in the process of making sense of the experiment and trying to find a model to describe it, he can keep in his mind at the same time all the variables $A, A', B,$ and $B'$. Then, to him, by a simple arithmetic argument, he first finds that the combination of variables $AB+A'B+AB'-A'B'=A(B+B')-A'(B-B')$ must take one of the values -2 or +2. ($B$ and $B’$ are either equal or opposite. In the first case $B-B’$ is zero, in the second case$ B+B’$ is zero. In both cases the remaining parenthesis is either -2 or +2. So, the whole expression is either -2 or +2.)

If Charlie is a mathematician or a statistician, he might try a joint probability model, and conclude from this and the observation above that the expectations E of the random variables satisfy $E(AB)+E(A'B)-E(AB')-E(A'B') \le 2$.
This is the most famous one of Bell's inequalities, the CHSH inequality. It can be violated by quantum mechanics, and all these experiments have shown that it also can be violated in the real world.

So, what is wrong with the simple argument above? I can see only one solution: Charlie is simply not able to all these 4 variables in his mind at the same time during his modeling efforts. Note that Charlie can be any person.
In my recent article [1] I have elaborated on such thoughts. The conclusions agree with the results of the Bell experiments, but they are not quite in agreement with what Tim Maudlin says.

Note that a hypothetical joint probability model for $A, A', B$ and $B'$ assumes that these are variables on some probability space, which in effect implies a hidden variable. I would say that in this sense the Nobel prize committee was right: The existence of such a hidden variable is excluded by the experimental result.

\section{Nonlocality and hidden varables?}

Tim Maudlin says that non-locality is the heart of the matter. Maybe so. But I have some difficulties in seeing this non-locality as so fundamental. Firstly, no real information can be passed from Bob to Alice. Secondly, it is always an assumption that the two particles had been in contact at some point of time in the past. If there were only one possible choice of measurement at each end, then this whole thing could have been understood by a version of what John Bell called a Bertlmann's sock's argument: Suppose that Bob always has two different socks, one red and one green. Suppose further that he at some point of time puts one of his socks in one box, the other in another box, and then sends a random box to Alice, who is far away. When then Alice opens her box and finds a red sock, she will immediately conclude that Bob has kept a green sock, and vice versa.

The strange thing by the Bell experiment, is that we may have a ‘common sense’ argument violated when both Alice and Bob have two possible maximal measurements, both socks and gloves at the same time in a way, the states of which are unknown both to Alice and to an independent observer Charlie.

Tim Maudlin refers to the theory of David Bohm in his arguments. I agree that this is a sort of hidden variable theory, and that it was well known to John Bell. However, as I read the press release of the Nobel prize committee, when they talk about hidden variables, they only mean this in the simple sense that I have mentioned above

\section{A new approach towards quantum foundation}
.
The observer Charlie may be any person, and his context may in general be a measurement process, a process related to an experiment, a model building process, or a process of trying to understand another person’s model. In short, any process during which he must make decisions.

I rely on the philosophy of the physicist Hervé Zwirn [5]. Zwirn says that every description of the world must be relative to the mind of an observer. I my discussions with Zwirn, I have claimed that also groups of communicating observers may have a joint description.
 
In my own last papers, I have tried to make Hervé' Zwirn’s philosophy concrete by proposing a simple model of our minds: Let us have variables $\theta ,\eta, \xi,...$  in our minds at some fixed time in some fixed context. Some of these may be accessible, that is, it is in principle possible to find values for them later by doing measurements or experiments. An accessible variable $\theta$ is called maximal if it cannot be extended to another accessible variable. 
And here is my main model assumption: There exists somehow an inaccessible variable $\phi$ such that all the accessible ones can be seen as functions of $\phi$.

Here are two important physical examples: 1) $\phi$ might be the vector (position, momentum) of a particle. 2) $\phi$ might be a spin vector of an electron. In other situations, the existence of a $\phi$ may be somewhat hypothetical.
Combining this simple model of the mind with some mathematically formulated symmetry-assumptions, I arrive at very interesting results. In [3] I derive very essential elements of quantum mechanics in this way. This is backed by arguments in the book [2]. Altogether, the whole of quantum mechanics can be derived from what I would call rather simple postulates.

This is a new and unfamiliar way of approaching quantum mechanics, but its basic postulates are not as formalistic as in the traditional approach. Results of the Bell experiment can be understood using my approach. Well-known paradoxes can by understood [3] using my approach. And the postulates can in principle be explained to any scientist with some mathematical background. Thus, the communication across scientific borders may be facilitated. 
\bigskip

\textbf{References}
\smallskip

[1] Helland, I.S. (2022a). The Bell experiment and the limitations of actors. Foundations of Physics 52, 55.

[2] Helland, I.S. (2021). Epistemic Processes. A Basis for Statistics and Quantum Theory. Springer, Berlin.

[3] Helland, I.S. (2022b). On reconstructing parts of quantum theory from two related maximal conceptual variables. International Journal of Theoretical Physics 61, 69.

[4] Maudlin, T. (2022). What the Nobel prize gets wrong about quantum mechanics. IAI news 6th October 2022.

[5] Zwirn, H. (2020). Nonlocality versus modified realism. Foundations of Physics 50, 1.

\end{document}